\documentclass[preprint,12pt]{elsarticle}

\usepackage{epsfig}
\usepackage{graphics}
\usepackage{graphicx}
\usepackage{hyperref}
\usepackage{amssymb}

\usepackage{lineno}

\usepackage{xspace}

\journal{Nucl. Instru. Meth. A}

\begin{document}

\def\Journal#1#2#3#4{{#1} {\bf #2}, #3 (#4)}

\def\NCA{Nuovo Cimento}
\def\NIM{Nucl. Instr. Meth.}
\def\NIMA{{Nucl. Instr. Meth.} A}
\def\NPB{{Nucl. Phys.} B}
\def\NPA{{Nucl. Phys.} A}
\def\PLB{{Phys. Lett.}  B}
\def\PRL{Phys. Rev. Lett.}
\def\PRC{{Phys. Rev.} C}
\def\PRD{{Phys. Rev.} D}
\def\ZPC{{Z. Phys.} C}
\def\JPG{{J. Phys.} G}
\def\CPC{Comput. Phys. Commun.}
\def\EPJ{{Eur. Phys. J.} C}
\def\PR{Phys. Rept.}
\def\JHEP{JHEP}

\def\pt{$p_{\rm T}$\xspace}
\def\dedx{$dE/dx$\xspace}

\begin{frontmatter}



\title{Calibration and performance of the STAR Muon Telescope Detector using cosmic rays}


\author[2]{C.~Yang}
\author[3]{X.J.~Huang\corref{cor1}}
\cortext[cor1]{Corresponding author, email address: huangxj12@mails.tsinghua.edu.cn}
\author[4]{C.M.~Du}
\author[1]{B.C.~Huang}

\author[8]{Z.~Ahammed}
\author[8]{A.~Banerjee}
\author[7]{P.~Bhattarari}
\author[8]{S.~Biswas}
\author[7]{B.~Bowen}
\author[5]{J.~Butterworth}
\author[6]{M.~Calder\'on de la Barca S\'anchez}
\author[9]{H.~Carson}
\author[8]{S.~Chattopadhyay}
\author[6]{D.~Cebra}
\author[2]{H.F.~Chen}
\author[3]{J.P.~Cheng}
\author[7]{M.~Codrington}
\author[5]{G.~Eppley}
\author[6]{C.~Flores}
\author[5]{F.~Geurts}
\author[7]{G.W.~Hoffmann}
\author[7]{A.~Jentsch}
\author[6]{A.~Kesich}
\author[2]{C.~Li}
\author[3]{Y.J.~Li}
\author[5]{W.J.~Llope}
\author[9]{S.~Mioduszewski}
\author[9]{Y.~Mohamed}
\author[5]{T.~Nussbaum}
\author[8]{A.~Roy}
\author[1]{L.~Ruan}
\author[7]{J.J.~Schambach}
\author[2]{Y.J.~Sun}
\author[3]{Y.~Wang}
\author[5]{K.~Xin}
\author[1]{Z.~Xu}
\author[2]{S.~Yang} 
\author[3]{X.L.~Zhu}
\address[1]{Brookhaven National Laboratory, Upton, New York 11973}
\address[2]{University of Science and Technology of China, Hefei 230026, China}
\address[3]{Tsinghua University, Beijing 100084, China}
\address[4]{Institute of Modern Physics, Lanzhou 730000, China}
\address[5]{Rice University, Houston, Texas 77005}
\address[6]{University of California, Davis, California 95616}
\address[7]{University of Texas at Austin, Austin, Texas 78712}
\address[8]{Variable Energy Cyclotron Centre, West Bengal 700064, India}
\address[9]{Texas A\&M University, College Station, Texas 77843}
\begin{abstract}
We report the timing and spatial resolution from the Muon
Telescope Detector (MTD) installed in the STAR experiment at RHIC. 
Cosmic ray muons traversing the STAR detector have an average 
transverse momentum of 6 GeV/$c$. Due to their very small multiple 
scattering, these cosmic muons provide an ideal tool to calibrate 
the detectors and measure their timing and spatial resolution. 
The values obtained were $\sim$100 ps and $\sim$1-2 cm, respectively.
These values are comparable to those obtained from cosmic-ray bench tests
and test beams.
\end{abstract}

\end{frontmatter}


\section{Introduction}\label{intro}
Data taken over the last decade have demonstrated that RHIC has
created a hot, dense medium with partonic degrees of freedom called
the Quark-Gluon Plasma (QGP). One
of the physics goals for the next decade is to study the
fundamental properties of the QGP such as the temperature, density profile,
and color-screening length via electro-magnetic probes such as
di-leptons~\cite{starwhitepaper,rhicwhitepaper,dilepton,dileptonII,
highptjpsi,satz_0512217,rhicIIQuarkonia,colorscreen,petreczky}.
Muons can be measured more precisely because of their relatively 
reduced Bremsstrahlung radiation in the detector materials. Such
an improved measurement is essential for separating the $\Upsilon$ meson
ground state (1S) from its excited states (2S+3S), each of which 
is predicted to melt at very different temperatures. The Muon
Telescope Detector (MTD) in the Solenoidal Tracker at RHIC (STAR)
will allow the measurement of the $\Upsilon$ mesons and $J/\psi$ mesons,
over a broad transverse momentum range through di-muon decays to
study color screening features, and $\mu$-e correlations to
distinguish heavy flavor correlations from initial lepton pair
production~\cite{starmtdproposal}. The MTD will thus provide direct
information on the temperature and the characteristics of 
color screening in the QGP created in RHIC collisions. 

The MTD is based on the Multi-gap Resistive Plate Chambers (MRPC) 
technology. A similar technology was used for the recently installed 
STAR TOF system~\cite{startofproposal,startof}. Unlike the TOF 
MRPCs~\cite{startofmrpcs}, however, the MTD MRPCs are much larger, and 
have long double-ended read-out strips. The MTD detectors are positioned
behind the iron return bars of the STAR magnet, and cover 45\% of the full 
azimuth within a pseudo-rapidity range of $|\eta|$$<$0.5.
The construction and installation of the MTD system was begun in 2011 
and will be completed in 2014.

Prototype MTD detectors were built and studied from 2007 to 2011. These
detectors were tested in the laboratory with cosmic rays, in test beams,
and in STAR experiment during RHIC runs.
The cosmic-ray and beam tests indicated a timing ($\le$100 ps) and the spatial 
resolution ($\sim$1 cm) that would be sufficient to achieve the physics
goals~\cite{MTDNIMA}. The operation of the prototype MTD detectors 
in STAR in 2007-2008 demonstrated that clean muon identification
could be achieved for muon transverse momenta above a few GeV/$c$. 
The bench and test-beam results, as well as detailed simulations, thus indicated
that the MTD would provide important physics information on
quarkonia and primordial di-lepton measurements at
RHIC~\cite{MTDPerformanceAtSTAR}. However, during the tests of the
MTD prototypes in STAR in 2007-2008, the timing resolution was 200-300 ps, 
which was much worse than that observed in the previous cosmic-ray and test-beam
studies. This poorer resolution resulted from the particular digitization
electronics~\cite{startrigger} that were used at that time
and the long cables between the MTD detectors and the digitizers. 

In 2009, new prototype MTD detectors were installed in STAR. For these prototypes,
the simple on-board front-end electronics used previously \cite{mtdelectronics} to drive
long cables to the digitizers were replaced with the same electronics as are used in the STAR TOF
system~\cite{startof,tofelectronics}. The TOF electronics are based on the CERN HPTDC~\cite{HPTDC} chip. 
In 2010, a cosmic-ray trigger based on the information from the MTD prototypes was implemented 
in the STAR trigger system. High energy ($\sim$6 GeV) cosmic rays 
provide an excellent means to study the timing and spatial resolution 
as the smearing from multiple scattering in the detector materials is a relatively small 
effect.

In this paper, the timing and spatial resolution of the MTD MRPCs read-out by 
STAR TOF electronics for cosmic muons reconstructed in the STAR experiment in 
2010-2011 will be reported.
The performance observed during the operation of 10\% of the full MTD system during
the 2012 RHIC run will also be presented.
This paper is arranged as follows. Section~\ref{mrpc} describes the MTD MRPCs, and 
Section~\ref{expr} describes the experimental aspects. The details of the data analysis and the
MRPC performance are reported in Section~\ref{results}. The summary and conclusions are then
presented in Section~\ref{concl}.

\section{MTD MRPC Modules}\label{mrpc}

A number of different MRPC designs were studied for possible use in the
MTD system. These include single- and double-stack MRPCs with five or six
gas gaps and several different read-out pad geometries. During the
2007-2011 RHIC runs, some of these prototype MRPCs were installed 
in STAR near mid-rapidity and at a radius of $\sim$400 cm. These prototypes 
were integrated into the STAR data acquisition and trigger systems and took
data throughout an entire RHIC run. Events useful for the study of the performance
of these detectors were saved using a trigger that required a hit in at least 
one read-out strip. 

\begin{table}[htbp]
\begin{center}
\caption{\label{Tab:RandD}The design parameters for the MTD MRPCs operated in STAR during
the 2007-2012 RHIC runs.}
\vspace*{1mm}
\begin{tabular}{|l|c|c|} \hline
Conditions & Type & trays \\ \hline 
cosmic and test-beam & type A & -- \\ \hline
Run 7: Au+Au &type A&2 modules in 1 tray  \\ 
Run 8: d+Au, p+p && \\ \hline
Run 9: p+p &type A&3 modules in 1 tray  \\ 
Run 10: Au+Au, cosmic trigger && \\ \hline
Run 11: Au+Au, cosmic trigger &type A&3 A modules in 1 tray \\ 
  & type B& 1 B module in 1 tray \\ 
  & type C& 2 C modules in 2 trays  \\ \hline
Run 12: p+p, cosmic trigger &type B&12 modules in 12 trays  \\  \hline
\end{tabular}
\end{center}
\end{table}

\begin{figure}[htbp]
\begin{center}
\includegraphics[keepaspectratio,width=0.9\textwidth]{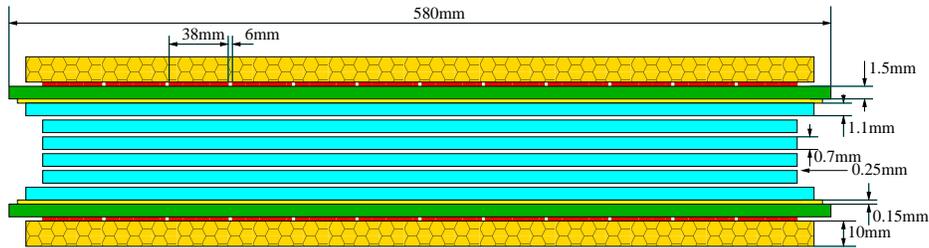}
\vspace*{-4mm}
\caption{(Color online) Side
view of a (type B) MRPC module, honey combs are colored yellow, 
strips are colored red, PCboards are colored green, mylars are 
colored light yellow and glasses are colored sky blue. The 
vertical scale has been expanded for clarity.} \label{mtd_lmrpc}
\end{center}
\end{figure}

Table~\ref{Tab:RandD} summarizes the details of the MRPC modules studied in the 2007-2012
RHIC runs. The MRPC modules were held in gas-tight aluminum boxes called ``trays."
A type A MRPC was a double-stack module with 10 gas gaps and 6 readout strips with outer 
dimensions of 87$\times$17 cm$^2$. The read-out pads were 
double-ended and 2.5 cm wide with 0.4 cm gaps in between each pad.
Type B (C) was a single-stack module with 5 (6) 250 $\mu$m-wide gas gaps and 12 
readout strips, also read out on both ends. The module size was 87$\times$52 cm$^2$, 
and the readout pads were 3.8 cm wide with 0.6 cm gaps. 
For the 2011 RHIC run, one MRPC of type B and two MRPCs of type C were installed in STAR.
The type B MRPC, shown schematically in Fig.~\ref{mtd_lmrpc}, was selected as the final 
design for the full MTD system. Twelve of these final-design modules were installed 
and operated during RHIC Run 12. 

The gas mixture used was 95\% Freon R-134a and 5\% isobutane. 
The readout electronics were very similar to those implemented in the
STAR TOF system. These record the absolute time of particle hits with
respect to a master 40 MHz clock with a time conversion for the 
least significant bit of 24.4 ps. The electronics also provide analog
multiplicity information from each end of each MRPC module (as an ``or"
of all of the strips in that module) to the STAR
Level-0 trigger. This information is used both for selecting events
with MTD hits, but also to enable triggers based only on prompt
hits, which will suppress the MTD backgrounds resulting from hadronic
showers in the magnet backlegs. Additional details on the 
electronics are available in Refs. \cite{startof,tofelectronics}.


\section{Experimental Setup}\label{expr}

\subsection{The STAR Detector}

The primary charged-particle tracking device
in the STAR experiment~\cite{star} is the Time
Projection Chamber (TPC)~\cite{startpc}. The TPC is 4 meters long
along the beam line and covers the full azimuth and a pseudo-rapidity range of $|\eta|$$<$1. 
Ionization electrons drift toward the nearest end, where the signals are read out. 
The TPC is divided into 24 super-sectors, 12 at each end, and each with 45 rows
of read-out pads along the radius. Each super-sector is divided
into inner and outer sectors. The TPC provides the positions of primary ionization
clusters generated by the passage of charged particles. The charged-particle tracks 
traversing the TPC will curve in the 0.5 T STAR magnetic field with a curvature 
that depends on the transverse momentum of the track. The ionization energy 
loss (\dedx ) measured as the areas of these charge clusters can be used for particle identification~\cite{pidNIMA,bichsel,pidpp08}. 

The STAR TOF detector \cite{startof} was fully installed before the 2010 RHIC run.
It covers the full azimuth and pseudo-rapidities $|\eta|$$<$0.9 at a radius of
$\sim$220 cm. It extends STAR's particle identification capabilities to momenta 
of $\sim$3 GeV/$c$ for $p$ and $\bar{p}$~\cite{tofPID}. The Barrel 
Electromagnetic Calorimeter (BEMC) is installed outside the TOF system~\cite{starbemc}. 
The TPC, TOF, and BEMC systems are centered in a solenoidal magnetic field 
generated by the STAR magnet. This magnet is cylindrical
and consists of 30 flux-return bars, four end rings, and
two pole tips. The return flux path for the field is 
provided by the outer magnet steel~\cite{starmagnet}. 
The 6.85 m long flux-return bars, also called ``backlegs," are trapezoidal in
cross-section and 57 cm thick at a radius of 363 cm.
These return bars and the BEMC serve as a hadron absorber allowing only muons 
to reach the MTD in elementary and heavy-ion collisions. 
Outside these return bars, there are the boxes which contain the photomultiplier
tubes (PMT) and electronics for the BEMC. The MTD detectors are mounted to the 
BEMC PMT boxes at a radius of $\sim$400 cm.

\subsection{STAR Cosmic Ray Trigger}

The simulation of a central Au$+$Au
collisions using a realistic description of the STAR geometry demonstrated that
most charged hadrons are stopped within the BEMC and backlegs and the few
escaping particles (primary or secondary) primarily pass through the
gaps between the backlegs~\cite{MTDPerformanceAtSTAR}. In the experimental
data collected during a RHIC run, the average transverse momentum, \pt, of muons
reaching the MTD detectors is $\sim$2 GeV/$c$. The multiple scattering for 2 GeV/$c$
muons in the backlegs leads to an MTD spatial resolution of 10 cm and a timing 
resolution of 100 ps. Multiple scattering thus makes it difficult to measure 
the timing and spatial resolution of the MTD MRPCs. Furthermore, with such 
a large smearing of the particle paths due to multiple scattering, the random 
matching between charged particles reconstructed in the TPC and MTD hits leads to 
significant backgrounds for identifying muons.

In the 2010 RHIC run, a cosmic-ray trigger based on the MTD detectors
was implemented. The desired cosmic
ray muons traverse the MTD, return bars, BEMC, TOF, and TPC detectors, 
generating long tracks in the TPC and two hits in the MTD and TOF barrels. 
As there was only a small patch of MTD detectors, and the TOF was fully installed,
the cosmic-ray trigger requires a coincidence of one MTD hit and two TOF hits. 
Hits belonging to a single muon traversing the TPC are collected and reconstructed
into tracks with well-defined trajectory, momentum $p$, and TPC ionization energy loss \dedx.
The STAR track reconstruction software assumes that particles originate in the center of 
STAR. Thus, a single cosmic muon passing through all of STAR is reconstructed as two
TPC tracks.

\begin{figure}[htbp]
\begin{center}
\includegraphics[keepaspectratio,width=0.45\textwidth]{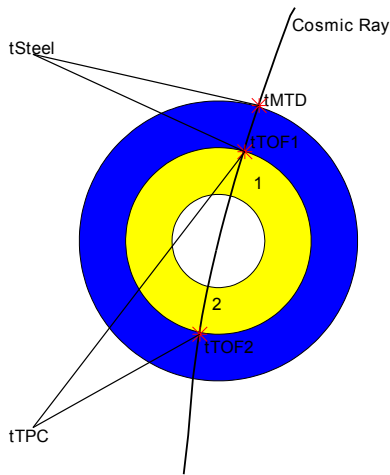}
\vspace*{-4mm}
\caption{(Color online) A schematic view of a cosmic-ray
event in the STAR detector.} \label{cosmicsketch}
\end{center}
\end{figure}

Figure~\ref{cosmicsketch} provides a schematic
view of a cosmic-ray event in the STAR detector. The time measured by the two 
TOF detectors and the MTD is called tTOF1, tTOF2, and tMTD, respectively. 
The calculated time of flight between the two TOF detectors, using the 
path length and $p$ measured by the TPC, is called tTPC. The time of flight 
from the MTD to the first TOF detector is called tSteel, which can be derived from 
the track's helix parameters, momentum $p$, and the 0.5 T magnetic field. 

\section{Performance of the MTD}\label{results}

\subsection{Muon Identification}

\begin{figure}[htb]
\begin{center}
\includegraphics[keepaspectratio,width=0.45\textwidth]{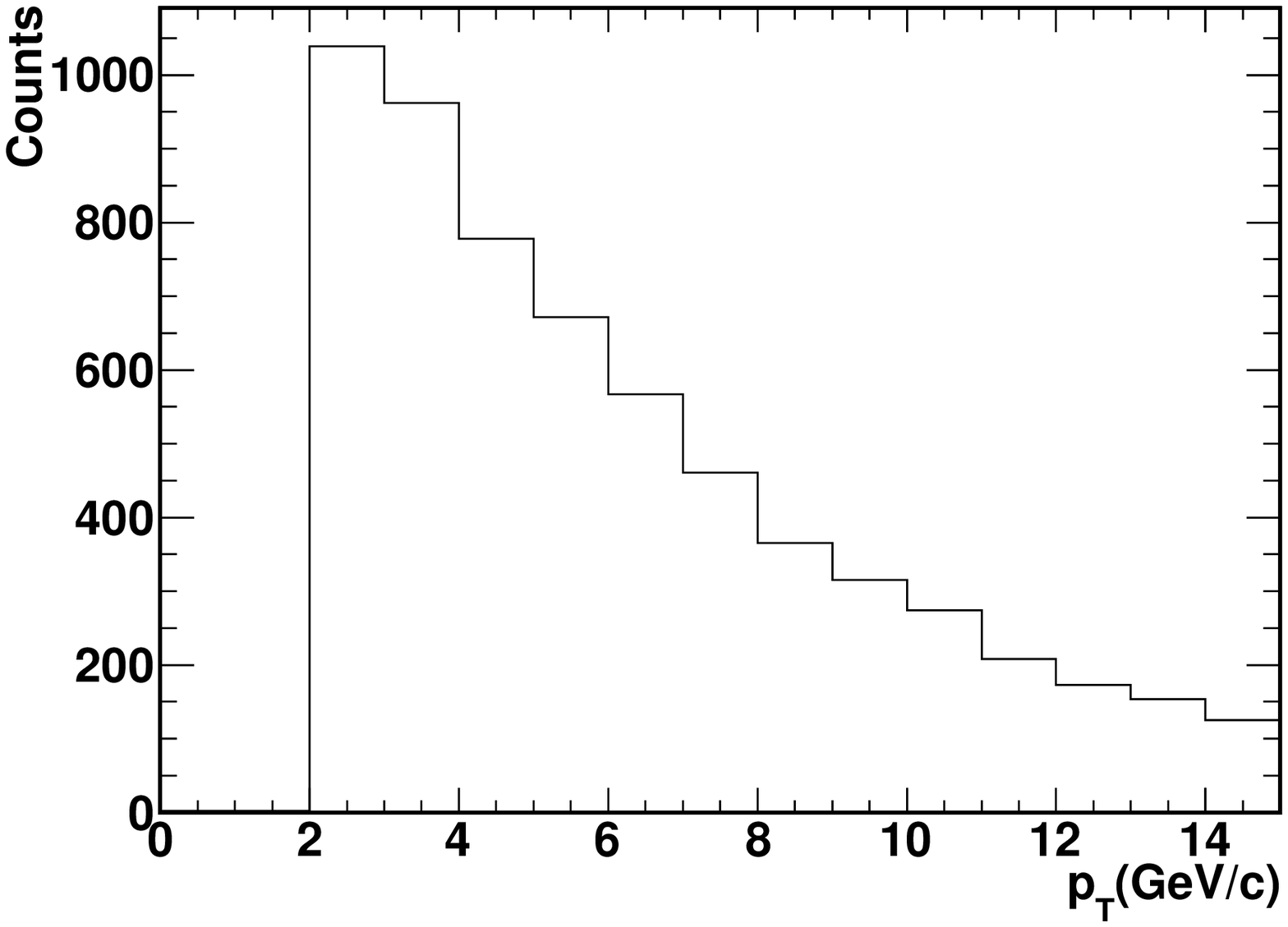}~~
\includegraphics[keepaspectratio,width=0.45\textwidth]{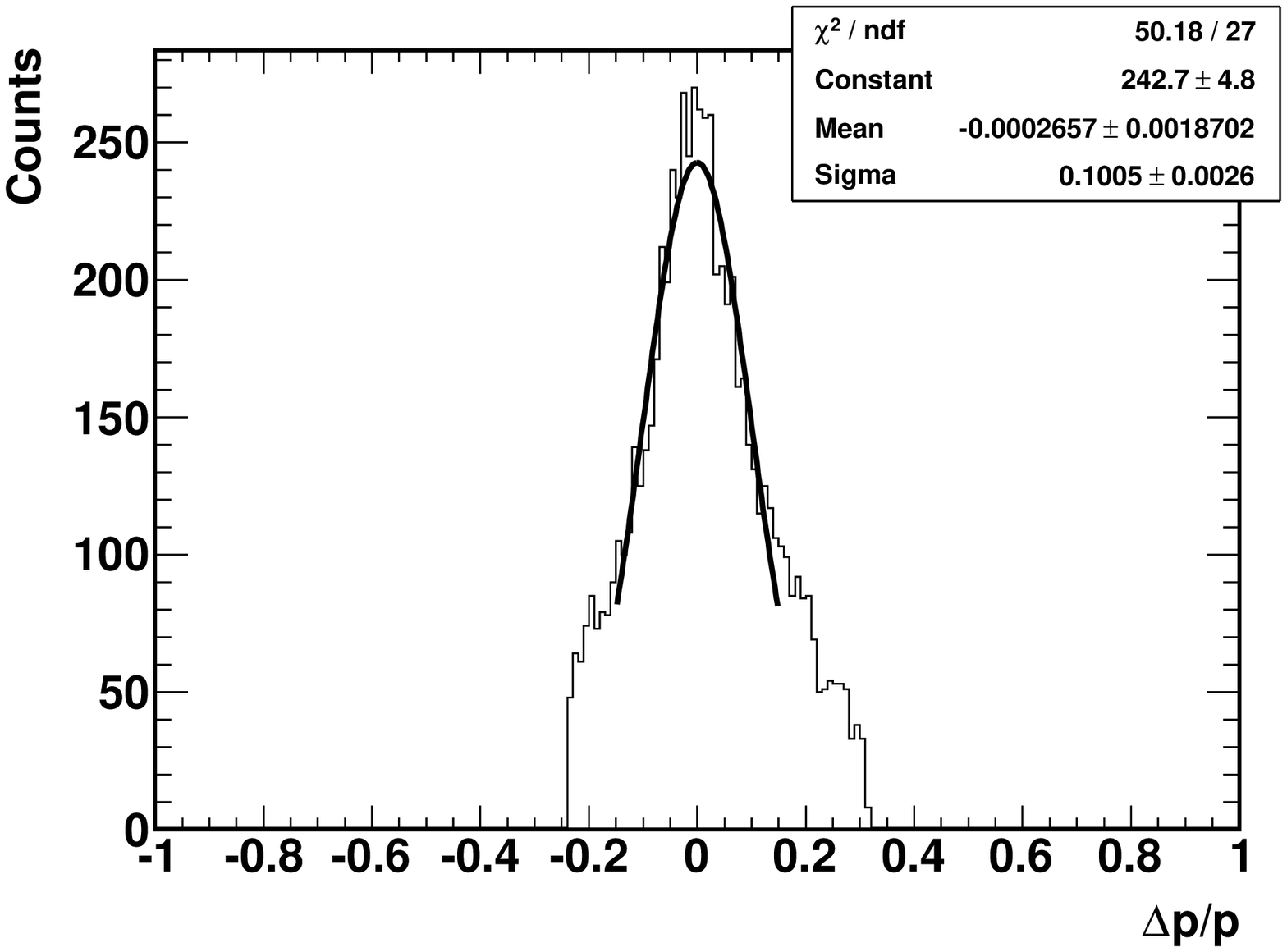}
\vspace*{-5mm}
\caption{ Left
panel: The \pt distribution of the cosmic-ray muons at \pt$\!>$2
GeV/$c$. Right panel: The relative momentum difference
distribution of the two tracks from one cosmic-ray muon. The curve
is a Gaussian fit to the distribution.} \label{ptres}
\end{center}
\end{figure}
\begin{figure}[htb]
\begin{center}
\includegraphics[keepaspectratio,width=0.7\textwidth]{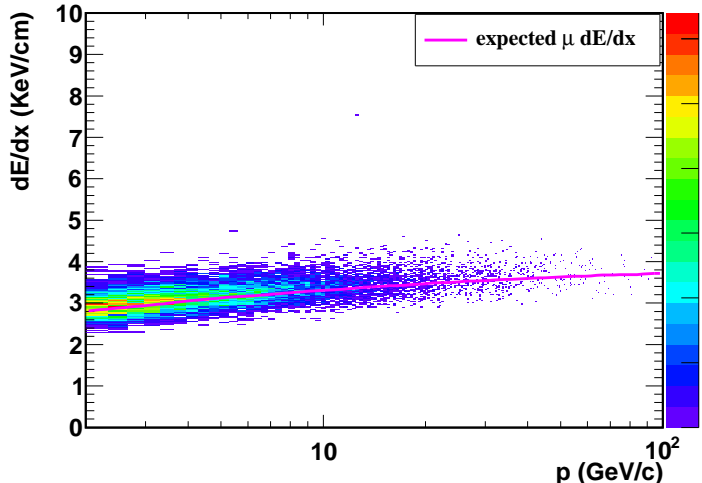}
\vspace*{-7mm}
\caption{(Color online) The TPC ionization energy loss,
\dedx, as a function of the momentum, $p$. The solid curve 
is the expected muon \dedx.} \label{dedx}
\end{center}
\end{figure}

Cosmic ray muon tracks with \pt$\!>$2 GeV/$c$ which matched to
TOF and MTD hits were selected. The two tracks from one cosmic
ray muon were each required to have at least 25 TPC hits (out of the
possible 45). The left frame of Fig.~\ref{ptres} shows the \pt 
distribution of the cosmic-ray muons with \pt$\!>$2 GeV/$c$, and the 
right frame shows the relative momentum difference
($\Delta p/p$) distribution of the two TPC-reconstructed tracks from one 
cosmic-ray muon. The average \pt is about 6 GeV/c and the average momentum
resolution is about 10\%. 

The absolute value of the relative momentum of the two TPC-reconstruct\-ed tracks, 
$\Delta p/p$, was required to be less than 27\%. The TPC ionization energy loss, \dedx, 
versus the momentum, $p$, is shown in Fig.~\ref{dedx}.
The expected muon \dedx versus $p$ from the Bichsel function is shown
as the line in this figure~\cite{bichsel}. It describes the data nicely, 
which demonstrates that a clean sample of cosmic muons was obtained.

\subsection{Spatial Resolution}

Tracks reconstructed in the STAR TPC were extrapolated to the 
MTD radius, which allows the measurement of the spatial
resolution of the MTD detectors in the $Z$ direction (along
the strips) and the azimuth, $\phi$, direction (perpendicular
to the strips). This extrapolation assumes a magnetic field
strength and direction that depends on the radius. The
magnetic field is 0.5 T in one direction for radii including
the TPC, TOF, and BEMC detectors, and is in the opposite direction
with a value of -1.26 T inside the backleg steel. The BEMC PMT
boxes and MTD detectors are in a field-free region.  

\begin{figure}[htbp]
\begin{center}
\includegraphics[keepaspectratio,width=0.48\textwidth]{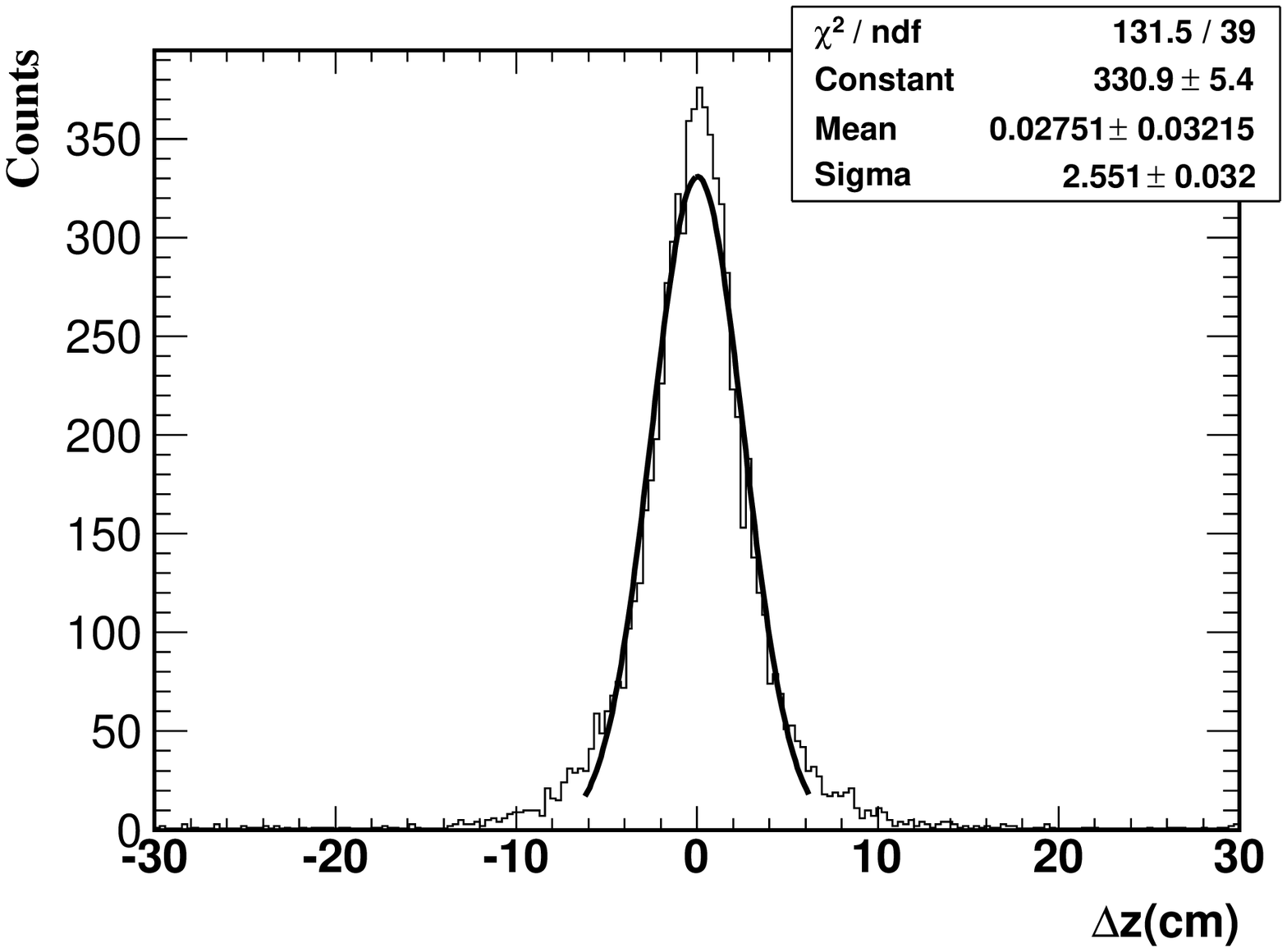}~
\includegraphics[keepaspectratio,width=0.48\textwidth]{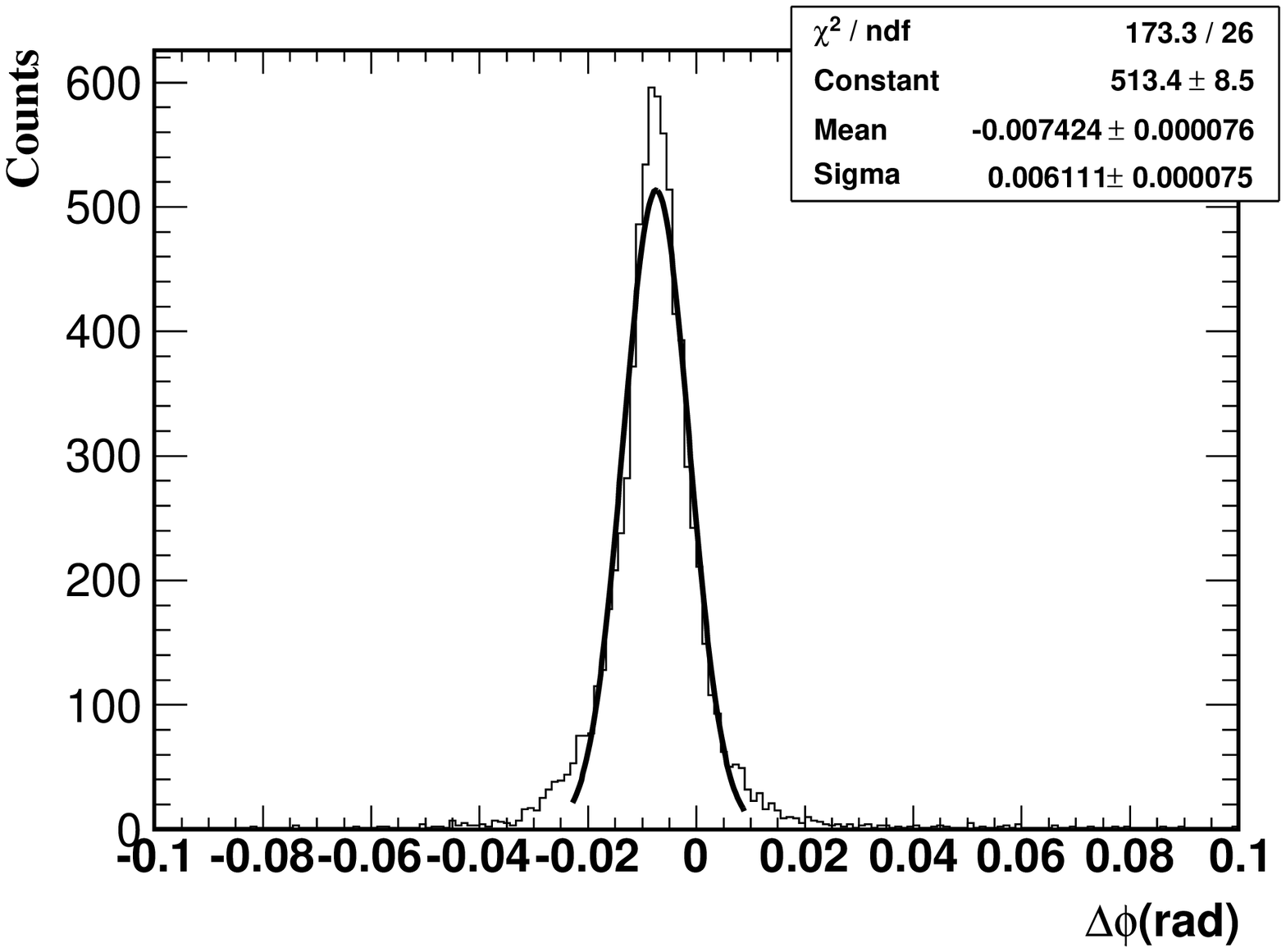}
\vspace*{-5mm}
\caption{ The $\Delta$$Z$ and $\Delta$$\phi$ distributions for the TPC-reconstructed
tracks extrapolated to the MTD detectors and the MTD-measured hits. The solid
curves are Gaussian fits to these distributions.}\label{zphireso}
\end{center}
\end{figure}

Figure~\ref{zphireso} shows the difference between the TPC track
extrapolation and the MTD hit position in the $Z$ direction, $\Delta$$Z$, (left frame) 
and in the $\phi$ direction, $\Delta$$\phi$, (right frame). For each MTD hit, 
the hit position in the $\phi$ direction is set at the middle of the strip with 
biggest signal, and the hit position in the $Z$ direction can be derived from the 
signal leading time difference between two readouts of the strip. The MTD detector
spatial resolution in the two directions are obtained as the standard deviations
of these difference distributions, and are 2.6 cm and 0.006 radians, respectively. 

\begin{figure}[htbp]
\begin{center}
\includegraphics[keepaspectratio,width=0.48\textwidth]{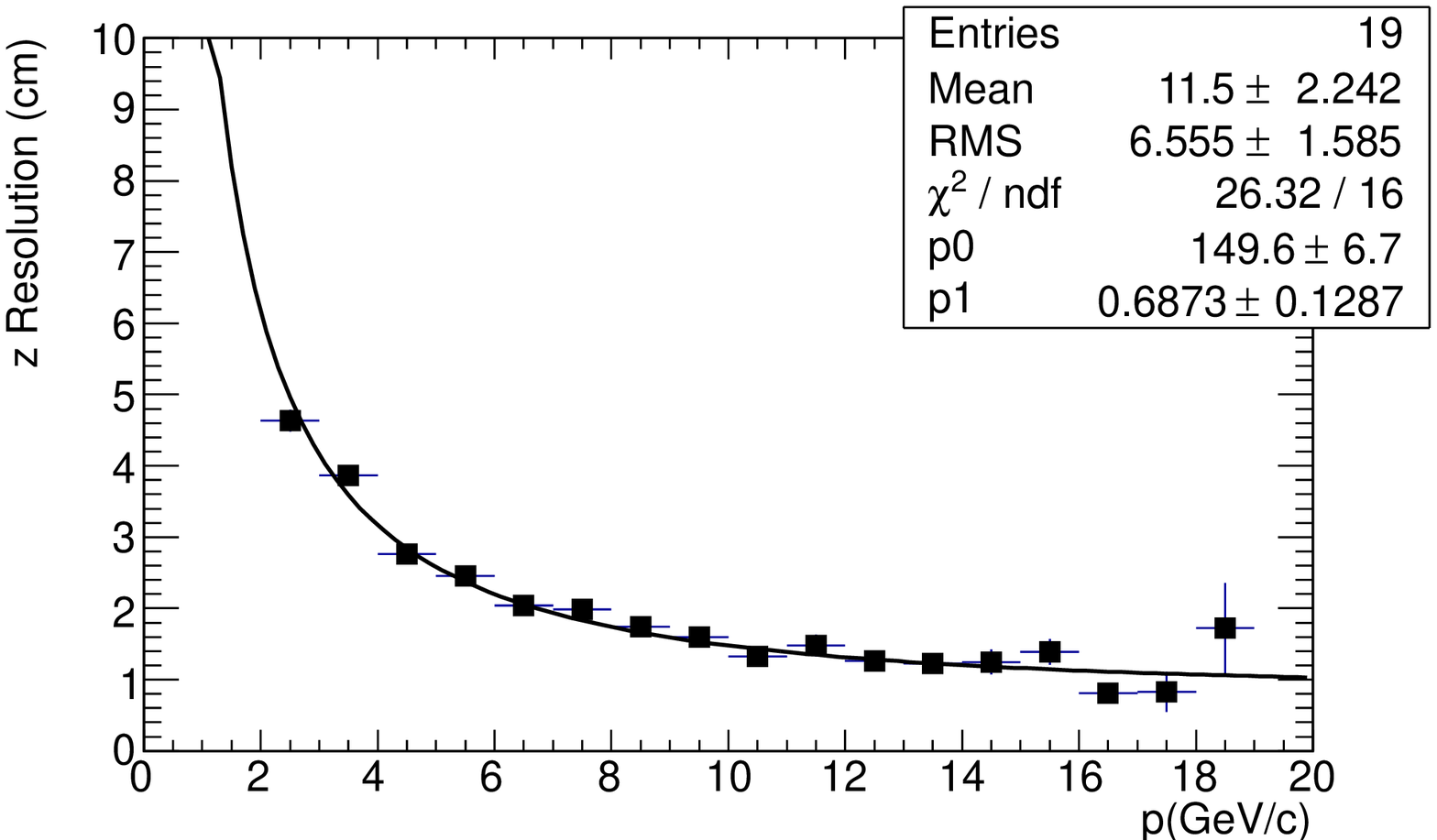}~
\includegraphics[keepaspectratio,width=0.48\textwidth]{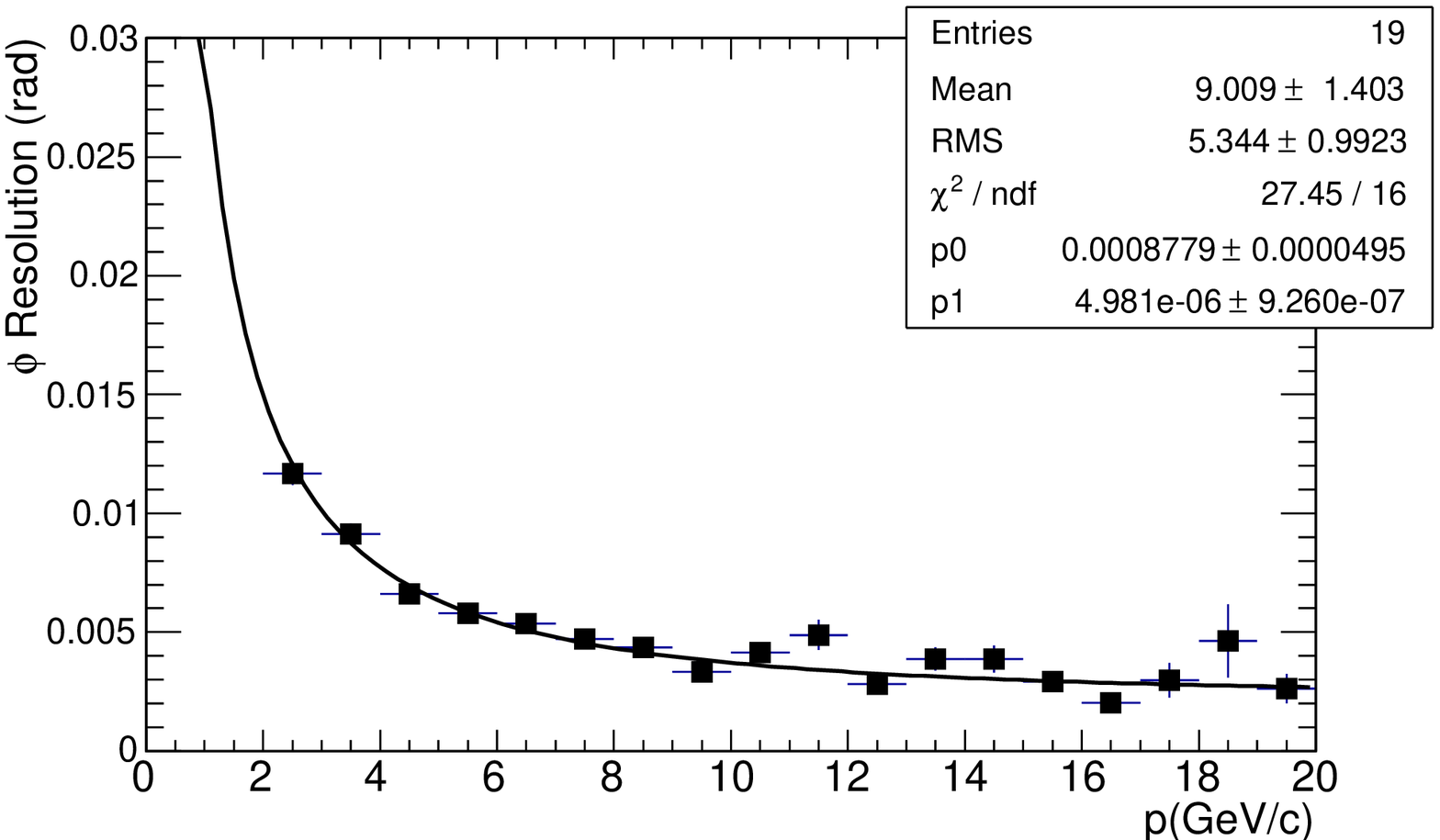}
\vspace*{-5mm}
\caption{The spatial resolution along the $Z$ (left frame) and
$\phi$ (right frame) directions as a function of the muon momentum.
The curves are fits to the data using the function described in the text.}\label{resovspt}
\end{center}
\end{figure}

Figure~\ref{resovspt} shows the dependence of the spatial resolution along the $Z$ (left frame)
and $\phi$ (right frame) direction as a function of the muon momentum. 
These data were fit with a functional form motivated by the expectation for the contribution
to the measured resolutions from multiple scattering in the detector materials. 
The formula used was $y$$=$$\sqrt{(p0/x^2)+p1}$, where $p0$ and $p1$ are fit parameters
and $x$ is the momentum. The spatial resolution of the MTD in the absence of multiple scattering in the
detector materials is then given by $\sqrt{p1}$. According to Fig.~\ref{resovspt},
the MTD detector spatial resolutions are then 0.8 cm and
2.2$\times$10$^{-3}$ rad in the $Z$ and $\phi$ directions, respectively.
The 2.2$\times$10$^{-3}$ rad $\phi$ resolution is equivalent to a 0.9 cm
resolution of the track at the 400 cm radius of the MTD detectors.

\subsection{Time Resolution}

The measurement of the time resolution of the MTD detectors is now described. 
Well-matched tracks were selected by requiring that $\Delta$Z$<$6 cm and 
$\Delta$$\phi$$<$0.2 rad. 

\begin{figure}[htbp]
\begin{center}
\includegraphics[keepaspectratio,width=0.7\textwidth]{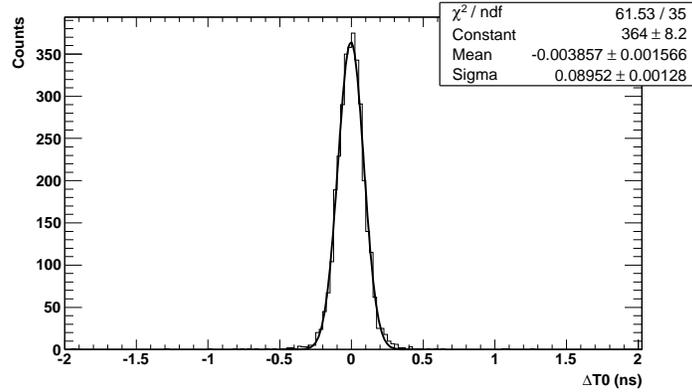}
\vspace*{-5mm}
\caption{The $\Delta$T0 distribution from the two TOF hits and the
expected time-of-flight obtained using the TPC-reconstructed track parameters. 
The curve is a Gaussian fit.} \label{tofreso}
\end{center}
\end{figure}

Two times for each muon track, called tTOF1 and tTOF2 ({\it cf.} Fig. \ref{cosmicsketch}),
were provided by the TOF detectors and are used to calculate the ``start time" for
the event. 
The resolution of the start time is obtained from the distribution of the time
difference, $\Delta$T0 $=$ (tTOF2$-$tTOF1)$-$tTPC. Here, (tTOF2$-$tTOF1) is
the TOF-measured time-of-flight, and tTPC is the time-of-flight that would
be expected for a muon with the trajectory and momentum as reconstructed
in the TPC. Figure~\ref{tofreso} shows the $\Delta$T0
distribution, from which a standard deviation of 90 ps is observed. The TOF
single-detector time resolution is thus 90 ps/$\sqrt{2}$, or 64 ps.

To perform the slewing calibration and evaluate the time resolution, the quantity of interest
is the MTD time, tMTD, minus the start time from the TOF detectors and the
expected time-of-flight for the cosmic muons between the TOF and MTD detectors, tSteel.
This is expressed as $\Delta$T $=$ (tTOF2$-$tTPC$+$tTOF1)/2 $-$ tMTD $-$ tSteel  
({\it cf.} Fig. \ref{cosmicsketch}).
The $\Delta$T distributions are recorded for each MTD read-out strip separately.

\begin{figure}[htbp]
\begin{center}
\includegraphics[keepaspectratio,width=0.45\textwidth]{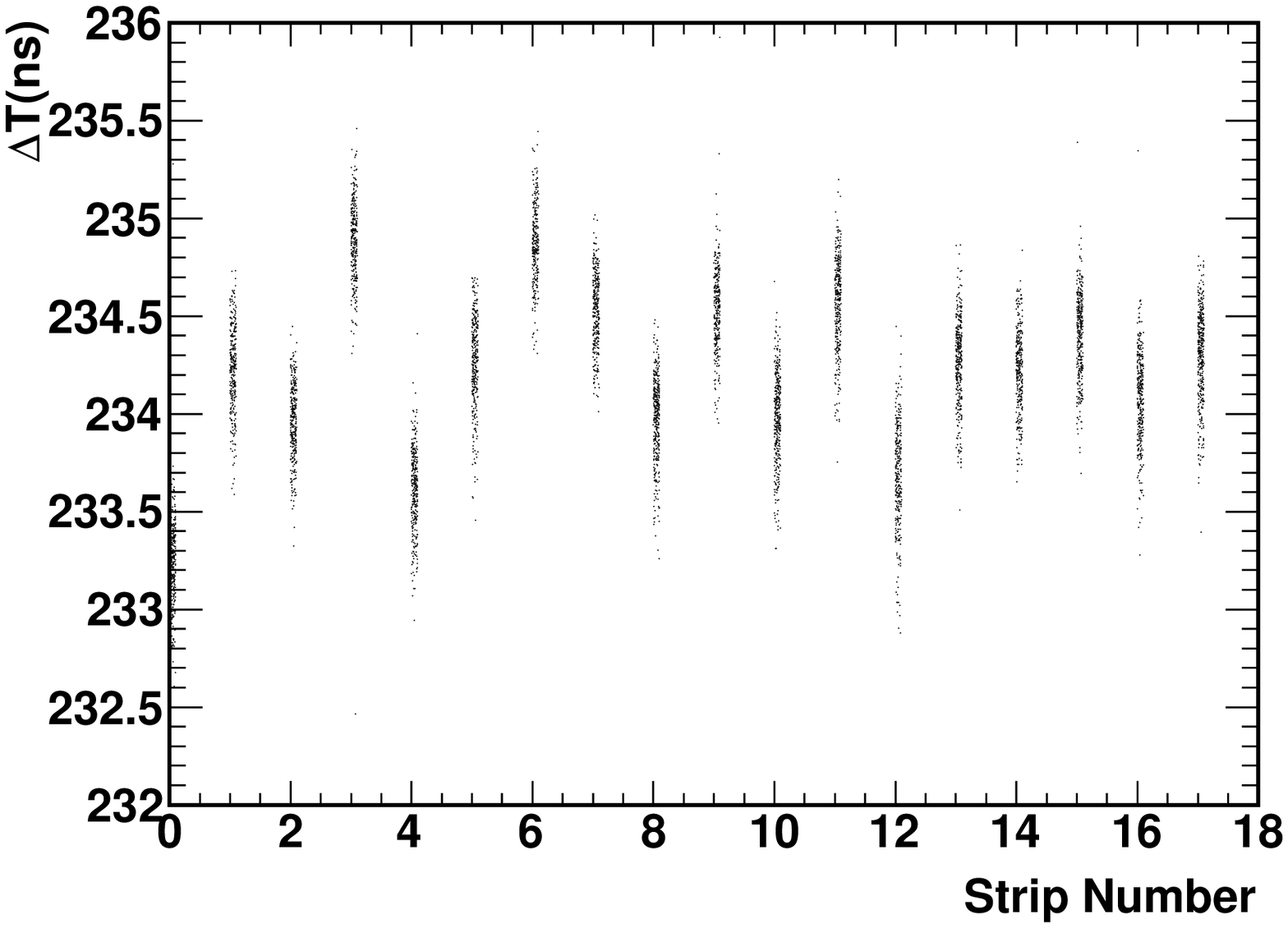}
\includegraphics[keepaspectratio,width=0.45\textwidth]{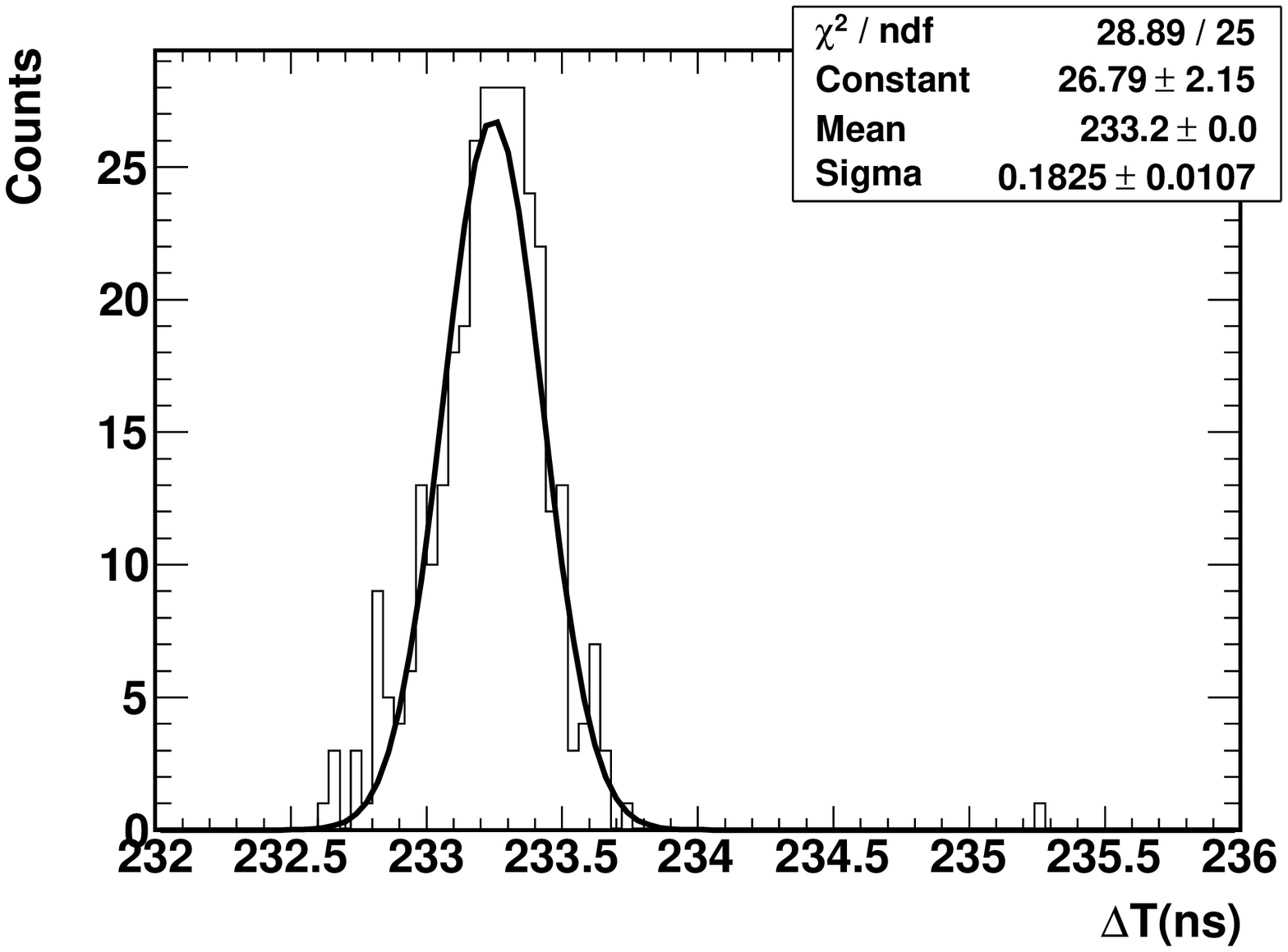}
\vspace*{-5mm}
\caption{The $\Delta$T distribution versus the readout strip (left frame), and
the $\Delta$T distribution and Gaussian fit from the first readout strip (right frame).} \label{mtdrawreso}
\end{center}
\end{figure}

The slewing calibration is done by three passes through the data. To illustrate
the procedure, the 2010 data with three type A modules will be used. 
In both the TOF and MTD systems, the pulse-size metric used for the
slewing correction is the pulse width at the leading-edge threshold, 
which is called ``Time over Threshold" (ToT). 

The first step is the extraction of the relative offsets, which proceeds by fitting the 
$\Delta$T distribution for each strip by a Gaussian, and recording the mean values of these 
Gaussians. The left frame of Fig.~\ref{mtdrawreso} shows  the $\Delta$T
distribution versus the readout strip, and the right frame shows 
the $\Delta$T distribution for one strip and the Gaussian fit. 

\begin{figure}[htbp]
\begin{center}
\includegraphics[keepaspectratio,width=0.7\textwidth]{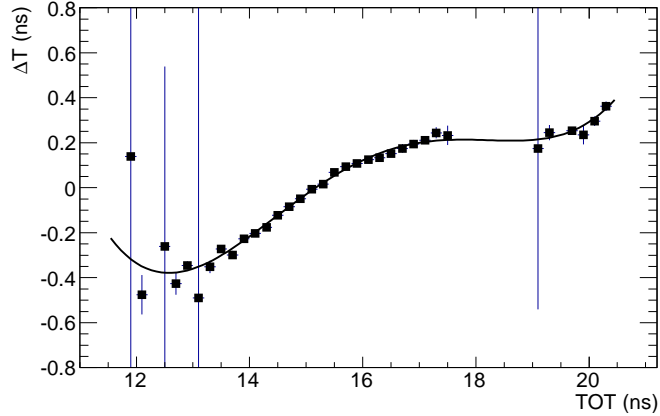}
\caption{The Gaussian mean $\Delta$T as a function of . The curve
is a fifth order polynomial fit to the data points.} \label{slewing}
\vspace*{-5mm}
\end{center}
\end{figure}

The second step to combine all channels using the relative offsets and 
plot the slewing itself, which is $\Delta$T versus
$\langle$ToT$\rangle$. The quantity $\langle$ToT$\rangle$ is the geometric
mean of the ToT values from the two ends of the lit strip, {\it i.e}, $\sqrt{{\rm ToT_L} * {\rm ToT_R}}$.
This is fit with a 5$^{th}$-order polynomial, and an example is shown in Fig.~\ref{slewing}.
The third step is re-equalize the channel-dependent time offsets after applying
the first step's crude offsets and subtracting the slewing using the 5$^{th}$-order polynomial.

\begin{figure}[htbp]
\begin{center}
\includegraphics[keepaspectratio,width=0.7\textwidth]{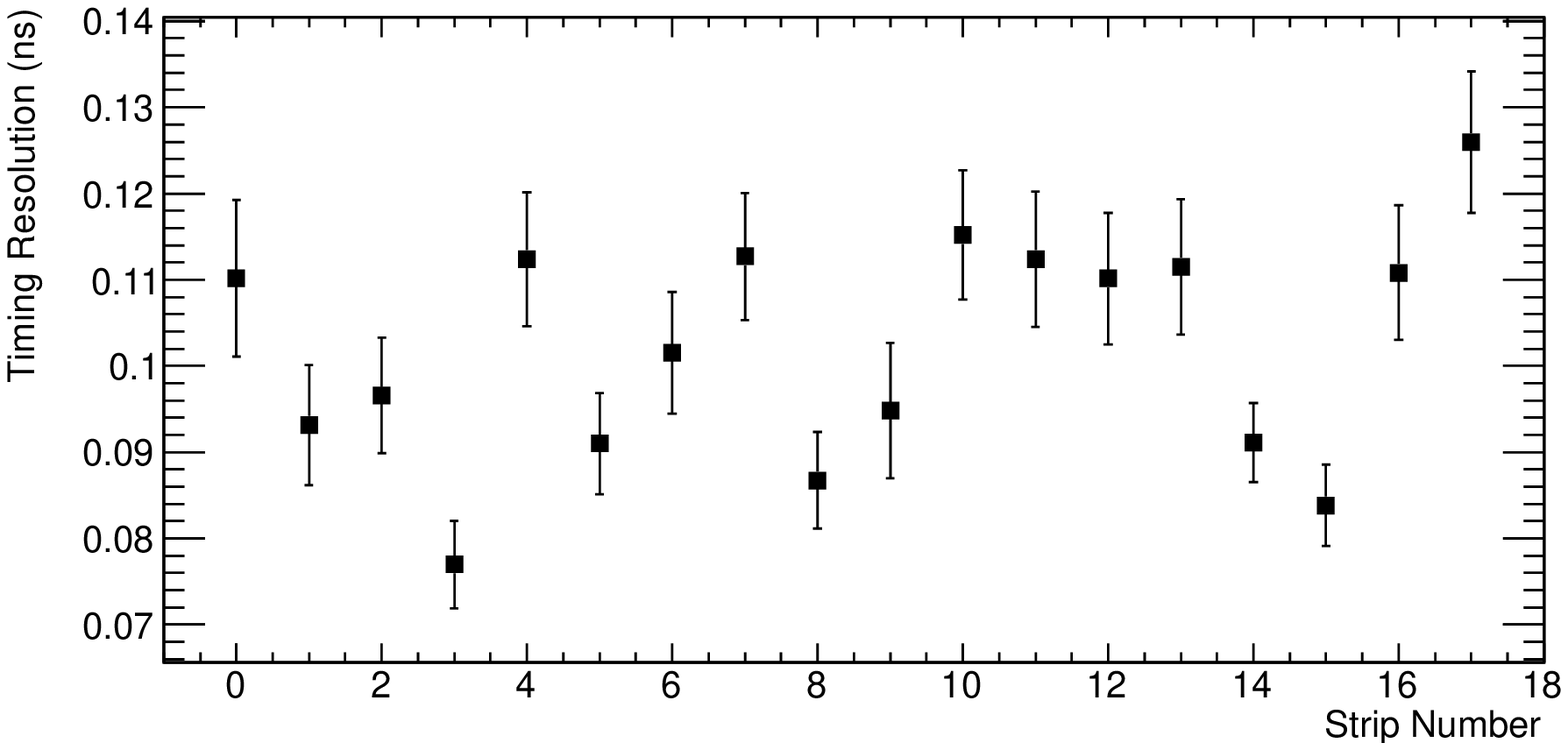}
\includegraphics[keepaspectratio,width=0.6\textwidth]{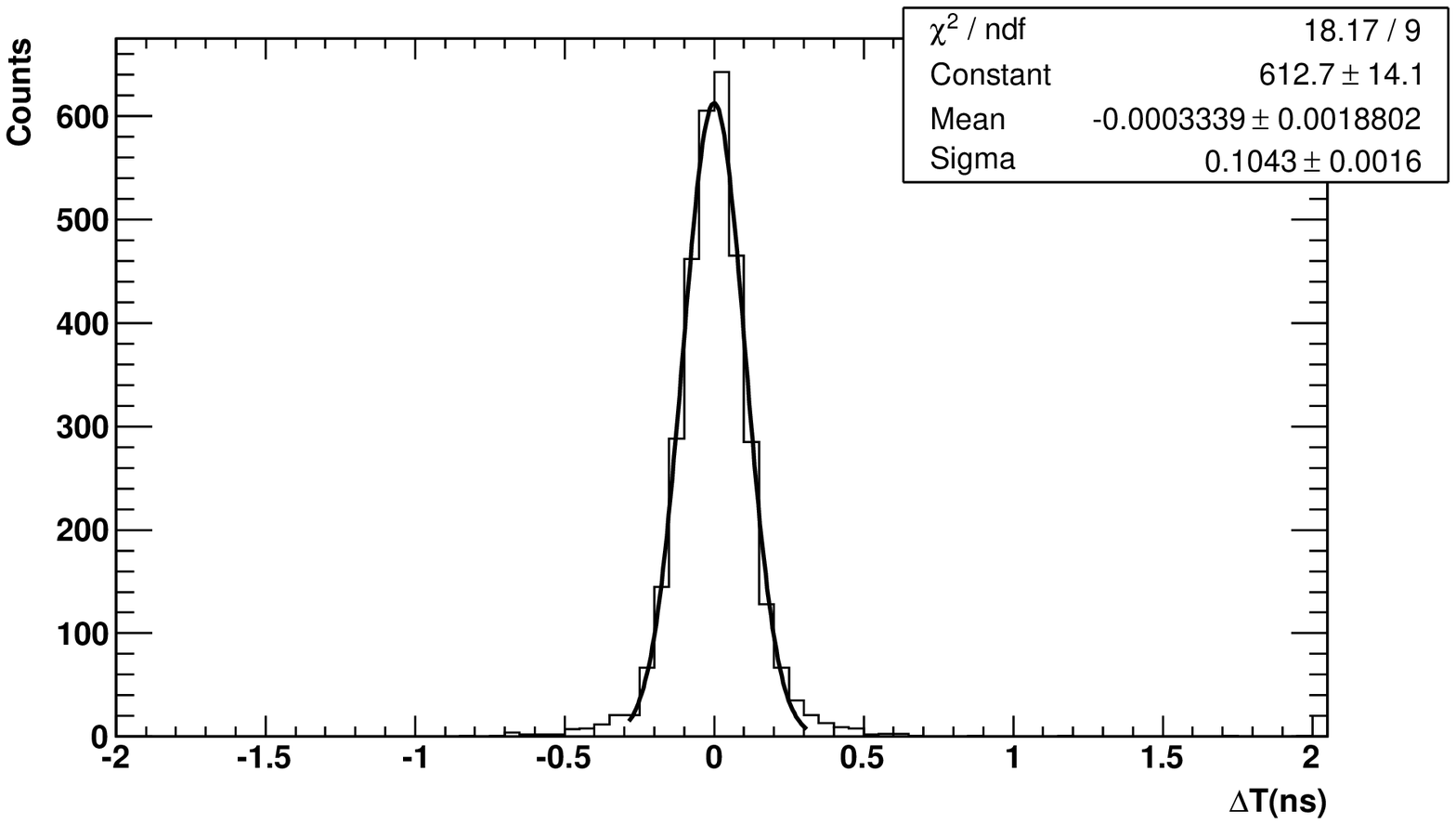}
\vspace*{-5mm}
\caption{The time resolution as a function of the readout channel (top frame), and 
the $\Delta$T distribution from all of the strips after the slewing and offset
corrections. The curve is a Gaussian fit to the distribution to obtain the total MTD+TOF timing resolution.}
\label{mtdfinalreso}
\end{center}
\end{figure}

The time resolution is then as the standard deviation of a Gaussian fit to the
resulting $\Delta$T distribution, which is shown for each strip in the top frame
of Fig.~\ref{mtdfinalreso}. Combining all of the strips and fitting that $\Delta$T distribution
with a Gaussian, which is shown in the bottom frame, implies a total time resolution of 104 ps. 
This resolution includes the start timing resolution of 90ps/2 $=$ 45 ps and the contribution
to the resolution from the multiple scattering for 6 GeV/$c$ muons, which is 25 ps.

After subtracting these two contributions in quadrature, the MTD time resolution is observed to be 90 ps.
This is close to the value obtained in the early test beam studies and the bench tests with cosmic rays.
The other data sets shown in Tab.~\ref{Tab:RandD} were calibrated using the same procedure and
the time and spatial position resolution results are summarized in Tab.~\ref{Tab:results}. 
The performance of the same type of MRPC detector in different years is consistent.  
 
\begin{table}[htbp]
\begin{center}
\begin{minipage}[t]{0.99\textwidth}
\begin{center}
\caption{The time and/or spatial resolution of the MTD MRPCs operated in STAR during
the 2007-2012 RHIC runs.\label{Tab:results}}
\vspace*{-4mm}
\begin{tabular}{|l|c|c|} \hline
Conditions & Type & resolution \\ \hline \hline 
cosmic \& test-beams~\cite{MTDNIMA} & type A & 60 ps, 0.6 cm \\ \hline
Run 7: Au+Au~\cite{MTDPerformanceAtSTAR} &type A & 200-300 ps  \\ 
Run 8: d+Au, p+p && \\ \hline
Run 9: p+p &type A & 90 ps, 0.8 cm ($z$), 0.9 cm ($\phi$)  \\ 
Run 10: Au+Au, cosmic trigger && \\ \hline
Run 11: Au+Au, cosmic trigger &type A & 89 ps, 1.1 cm ($z$), 1.1 cm ($\phi$)   \\ 
  & type B & 101 ps, 1.4 cm ($z$), 1.6 cm ($\phi$)\\ 
  & type C &  127 ps, 2.4 cm ($z$), 1.8 cm ($\phi$) \\ \hline
Run 12: p+p, cosmic trigger & type B& 108 ps, 2.6 cm ($z$), 1.9 cm ($\phi$)   \\   \hline
\end{tabular}
\end{center}
\end{minipage}
\end{center}
\end{table}

\section{Conclusions}\label{concl}

The time and spatial resolution of the STAR MTD system were obtained using cosmic-ray muons traversing the
STAR detector. The relatively high momentum muons can be cleanly triggered upon and selected in the
data, and allow studies of the time and spatial resolutions with relatively small contributions
from multiple scattering in the STAR detector materials. 
The MTD resolution values observed in data sets spanning several years were 100~ps for the timing resolution
and 1-2 cm for the spatial resolution.


\section{Acknowledgments}
We thank the STAR Collaboration and the RACF at BNL for their support. 


\end{document}